\documentclass[preprint,pre,aps,showpacs,amsmath,eqsecnum]{revtex4}
\usepackage{graphicx}
\usepackage{dcolumn}

\usepackage{bm}
\begin{document}
\title{Sharp-interface projection of a fluctuating phase-field model}
\author{R. Ben\'{\i}tez}
\author{L. Ram\'{\i}rez-Piscina} 

\affiliation{
Departament de F\'{\i}sica Aplicada,\\
Universitat Polit\`ecnica de Catalunya,\\
Doctor Mara\~n\'on 44, E-08028 Barcelona, Spain.
} 

\date{\today}

\begin{abstract}
We present a derivation of the sharp-interface limit of a generic fluctuating 
phase-field model for solidification. 
As a main result, we obtain a sharp-interface projection 
which presents noise terms in 
both the diffusion equation and in the moving boundary conditions.
The presented procedure does not rely on the fluctuation-dissipation theorem, 
and can therefore be applied to account for both internal and external 
fluctuations in either variational or non-variational phase-field formulations. 
In particular, it can be used to introduce 
thermodynamical fluctuations in non-variational formulations 
of the phase-field model, which permit to reach better computational 
efficiency and provide more flexibility for describing some features of specific 
physical situations. This opens the possibility of performing quantitative 
phase-field simulations in crystal growth while accounting for 
the proper fluctuations of the system. 
\end{abstract}

\maketitle

%
%
\section{INTRODUCTION}

%
%
In recent years, phase-field models
have emerged as an efficient technique to 
simulate interfacial phenomena in 
non-equilibrium systems \cite{ricardcuerno}. 
This method has mainly been developed for solidification \cite{WBM1,WBM2,karma-thin}, but has also successfully been applied to other problems, such as grain boundaries \cite{warren-grain-boundaries},
crack propagation \cite{crack-aranson}, viscous fingering \cite{Folch99} 
or vesicle dynamics \cite{vesicles2}. 
The phase-field approach introduces an 
equation for a continuous 
variable $\phi({\bf r},t)$, which appears as an order parameter, and takes 
distinct, constant values in the different phases.
The interface is then described by the level set $\phi=\text{\it constant}$, 
and the transition between both phases 
takes place in a diffuse interface of thickness $W$. 
The model is completed by coupling 
the $\phi$ equation with a diffusion field 
which acts as a driving force for the motion of the front.
The behavior of the diffuse interface
can then be computed by the integration of a set of partial differential equations
for the whole system, therefore avoiding the explicit tracking of the interface
position. 
This has practical advantages over using
the free boundary conditions that are characteristic 
of a moving boundary description. 
Phase-field models are usually constructed to 
recover the classical moving boundary dynamics 
in the so called sharp-interface limit as $W \rightarrow 0$ \cite{caginalp-fife}.
This limit is taken by means of a systematic 
asymptotic expansion on the interface width, and allows the model 
parameters to be determined in terms of the physical properties of the system. 

%
%
In early phase-field formulations, the model equations 
were derived from the variational minimization of a global free-energy 
functional for the heterogeneous system. Such variational formulations, 
however, in spite of their appealing structure, presented poor computational 
efficiency and did not permit to obtain truly quantitative results. 
For this reason, recently proposed phase-field models are 
not derived from a variational principle, but are specifically constructed 
to recover a certain moving boundary problem in the 
sharp-interface limit \cite{Folch99,elder-grant}. 
Besides presenting a better computational behavior, non-variational phase-field 
formulations provide for more flexibility in the description of 
some particular features such as different transport 
properties in the solid and liquid phases \cite{almgren-one-sided,karma-one-sided}. 

%
%
On the other hand, the presence of fluctuations has always been an 
important issue in the study of pattern-forming instabilities in crystal growth 
\cite{karma-fluctuations,karma-fluct-prl}. 
Indeed, internal or external noises play the role of an initiation mechanism for the 
morphological deformations of the interface \cite{wl-1,wl-2}. Thermal or solute 
fluctuations, for instance, must be taken into account in order to 
study important problems such as the dynamical selection of the primary spacing in 
directional solidification \cite{wl-2} or the formation of secondary instabilities 
(sidebranches) in dendritic growth \cite{langer_sidebranches}. 
%
%
Fluctuations were soon introduced into phase-field 
models in an {\it ad hoc} way as a controlled source of interfacial 
perturbations \cite{kobayashi}. However, phase field models accounting 
for internal thermodynamical fluctuations have not been proposed until recently, and in the context of variational formulations 
\cite{elder-eutectic-noise, karma-pf-noise,pavlik-sekerka-1,pavlik-sekerka-2}.  
In such variational cases, the statistical properties of the fluctuating terms can 
straightforwardly be determined by using the fluctuation-dissipation theorem, 
following the lines applied by Hohenberg and Halperin within the 
context of critical dynamics \cite{hohenberg-halperin}. 
In non-variational formulations, however, the fluctuation-dissipation relation 
becomes useless for this purpose because the dynamics of the system cannot 
be derived from a single free-energy functional. 

%
%
The aim of this work is to present a systematic procedure 
to account for the introduction of generic sources of noise in either 
variational or non-variational phase-field models. 
To this end, we will perform the sharp-interface limit of a fluctuating phase-field 
model for solidification and explicitely obtain 
the properties of the projected noise terms 
that will appear in the moving boundary equations.
This projection, which does not rely on the fluctuation-dissipation theorem,
will be carried out by means of a hybrid asymptotic expansion which 
combines a standard sharp-interface limit with a small noise assumption 
for the intensities of the noise terms in the model. 
The structure of the resulting sharp-interface projection takes the form 
of a moving boundary problem, which now includes 
bulk and interfacial stochastic terms. The statistical properties of 
these new terms are related to those of the noises appearing in 
the starting phase-field equations. The extension of our procedure to 
thin-interface asymptotics \cite{karma-thin} is straightforward 
and is not presented here for the sake of clarity.  

As a particular case, this analytical technique will enable a prescription 
for the introduction of internal thermodynamical fluctuations 
in non-variational phase-field models, subject only to the constraint 
of providing the correct interface equilibrium fluctuations. This approach will also
allow for the consideration of more general noise sources of an 
external origin, such as experimental imperfections or 
controlled perturbations, which do not follow equilibrium statistics. 
It is worth pointing out that while the calculations
will be performed within the framework of the symmetric solidification model, the 
approach can be easily extended to one-sided formulations \cite{karma-one-sided}.

This work has been organized as follows: 
The stochastic model equations are presented in 
Sec.~\ref{sec:model}. The asymptotic stochastic 
procedure is developed in Sec.~\ref{sec:hybrid}, which has been 
divided in four different subsections: Sec.~\ref{sec:a} 
and \ref{sec:b} are dedicated to find solutions of the equations in 
the inner and outer asymptotic regions, respectively. 
The solvability conditions for the inner expansion are imposed in 
Sec.~\ref{sec:c}, whereas in Sec.~\ref{sec:d} we perform the 
asymptotic matching between the inner and outer 
stochastic fields in order to obtain the form of the projected equations. 
The projected problem is then compared in Sec.~\ref{sec:parameters} with 
the standard Lanvegin formulation for solidification \cite{karma-fluctuations,karma-fluct-prl}, allowing for the 
determination of the model parameters in the case of having internal noises 
of a thermodynamical origin. A numerical test for the validity of the 
approach is reported in Sec.~\ref{sec:test}, and Sec.~\ref{sec:discuss} is devoted 
to present some discussion and concluding remarks. 
%
%
\section{MODEL EQUATIONS}
\label{sec:model}
Our approach starts from a generic 
non-variational phase-field model, which applies for both the 
solidification of a pure substance and for the symmetric solidification 
of a dilute alloy with a constant miscibility gap \cite{karma-thin},
\begin{eqnarray}
\alpha \varepsilon^2 \partial_{t} \phi &=& \varepsilon^2\nabla^2\phi-
f'(\phi)- \varepsilon \lambda  g'(\phi) u + \varepsilon^{\frac{3}{2}}\eta({\bf
r},t)
\label{eq:pf-1}\\
\partial_{t} u &=& \nabla^{2}u+\frac{1}{2}\partial_{t} h(\phi) -
{\bf \nabla}\cdot {\bf q}({\bf r},t)\;,
\label{eq:u-1}
\end{eqnarray}
where $\alpha$ is a parameter determining the 
time scale of the phase-field dynamics and $\lambda$ 
accounts for the coupling strength between $\phi$ and the diffusion 
field $u$. 
We choose $g(\phi)$ and $h(\phi)$ to be odd polynomial functions of $\phi$ 
satisfying the limiting conditions $g'(\pm 1)=0$ and $h(\pm 1)=\pm 1$, 
and $f(\phi)$ to be given by the 
standard double-well potential 
\begin{equation}
f(\phi)=\frac{1}{4}\phi^4-\frac{1}{2}\phi^2.
\label{eq:double-well}
\end{equation}
In the model equations Eqs.~(\ref{eq:pf-1}), (\ref{eq:u-1}), $u$ is a reduced diffusive field 
defined by $u=(T-T_M)/(L/c)$ in the case of pure substances and by 
$u = ( c - \frac{1}{2}(c^0_S+c^0_L))/\Delta c_0 + \frac{1}{2}g(\phi)$ 
for symmetric alloys, where $T_M$ is the melting temperature, 
$L$ the latent heat per unit volume, $c$ the specific heat per unit volume 
and $\Delta c_0 \equiv c^0_L-c^0_S$, being $c^0_S,c^0_L$ the solid and liquid 
equilibrium concentrations of the alloy, respectively.  
The two minima $\phi=\pm 1$ of $f(\phi)$ in Eq.~(\ref{eq:double-well}) correspond respectively to the solid and liquid phases of the system, so the interface will be represented by the transition zone between these two values.
Space and times in Eqs.~(\ref{eq:pf-1}), (\ref{eq:u-1}) have been scaled out using a characteristic length 
$l$ and a time scale $\nu=l^2/D$, where $D$ is the thermal or chemical 
diffusivity of the substance. The control parameter $\varepsilon=W/l$ is the scaled 
interface thickness, and will be the small parameter in which the formal
expansions will be carried out. 

Fluctuations appear in the model as a non-conserved noise term $\eta$ in 
the equation for the phase-field, and as a conserved stochastic current ${\bf q}$ 
in the diffusion equation. These fluctuations account 
for generic noise sources of either an internal 
or an external origin. 
We assume that the noises are white and Gaussian with correlations given by  
\begin{eqnarray}
\langle
\eta({\bf r},t) \eta({\bf r'},t') 
\rangle &=&
2 \sigma_\phi^2 \delta({\bf r}-{\bf r'}) 
\delta(t-t')\;,
\label{eq:eta}\\
\langle
q_i({\bf r},t)
q_j({\bf r'},t')
\rangle &=&
2 \sigma_u^2 \delta_{ij} \delta({\bf r}-{\bf r'})\delta(t-t')\;.
\label{eq:q} 
\end{eqnarray}
%
%
In the proposed phase field model,
parameters such as $\alpha$, $\lambda$, and the noise amplitudes $\sigma_\phi$, $\sigma_u$, are intended to represent (or to be directly related to) physical parameters. On the contrary, the scaled interface width $\varepsilon$ has been introduced as an expansion parameter. As a matter of fact, 
Eqs.~(\ref{eq:pf-1}), (\ref{eq:u-1}) have been constructed so that
the resulting dynamics (in the double limit of sharp interface and small noise) 
will be independent of $\varepsilon$.
In particular, the scaling factor $\epsilon^{3/2}$ of the the noise term in 
Eq.~(\ref{eq:pf-1}) has been introduced 
in order to make the fluctuations of the interfacial dynamics, as will be obtained below, 
independent of $\epsilon$. 
The details of this calculation and the presentation of the results are given in the next section. 
%
%
\section{HYBRID ASYMPTOTIC EXPANSION}
\label{sec:hybrid}

In order to deal with fluctuating phase-field models, 
the standard asymptotic expansion, 
performed in terms of a small interface thickness, 
should be complemented with a small noise assumption. 
The combination of these approaches will give rise to 
a hybrid asymptotic procedure.
To this end, the small noise assumption will be imposed by 
assuming that $\sigma_\phi, \sigma_u$ obey some order 
relations with the interface thickness $\varepsilon$. 
Namely, we take 
\begin{eqnarray} 
\sigma_\phi &\sim& O(\varepsilon^{3/2}),
\label{eq:sigma_order1} \\
\sigma_u &\sim& O(\varepsilon^2), 
\label{eq:sigma_order2}
\end{eqnarray} 
which will permit along the expansion procedure to maintain the fluctuating terms as small perturbations at the desired order in a consistent way.
Relations (\ref{eq:sigma_order1}), (\ref{eq:sigma_order2}) should not be understood as any explicit dependence of these parameters on $\varepsilon$, 
but only as a way to formalize a double expansion in terms of a 
single vanishing parameter, the interface thickness, $\varepsilon \rightarrow 0$. 
%
%
%

Our method closely follows the standard asymptotic procedure described in 
Ref.~\cite{almgren-one-sided}. 
We start by dividing the system into two different regions: 
an outer region far from the interface at distances much greater than 
$\varepsilon$, where the phase field presents the two constant values $\phi=\pm 1$ representing the solid and liquid phases at each side of the interface,  
and an inner region located around the interface up to distances of order 
$\varepsilon$, where the phase field varies between these two values. 
In the limit $\varepsilon \rightarrow 0$, solutions 
for the fields in both regions should match order by order 
in $\varepsilon$ at some intermediate distance $r_M$, which can be taken of 
order $r_M \sim \varepsilon ^{1/2}$.
%
%
\subsection{Outer region}
\label{sec:a}
%
%
In the outer region, the equations can be solved at each order by expanding 
the fields in powers of $\varepsilon$ as
\begin{eqnarray}
u &=& u_0+\varepsilon u_1+\varepsilon^2 u_2+O(\varepsilon^3),\\
\phi &=& \phi_0 + \varepsilon\phi_1+ \varepsilon^2\phi_2+O(\varepsilon^3),
\label{eq:outer}
\end{eqnarray}
and by expanding in Taylor series around $\phi=\phi_0$ the functions $f,g$ 
appearing in the model equations. 
%
If we use the order relations Eqs.~(\ref{eq:sigma_order1}), (\ref{eq:sigma_order2}), 
the noise terms can be assumed to be of orders 
\begin{eqnarray} 
\eta &\sim& O(\varepsilon^{3/2}),
\label{eq:eta_order1} \\
{\bf q} &\sim& O(\varepsilon^2), 
\label{eq:q_order2}
\end{eqnarray} 
and the outer equations can then be obtained at each order in $\varepsilon$.
\subsubsection{Zero Order}
At the leading order ($\varepsilon^0$), the outer equations are given by
\begin{eqnarray}
f'(\phi_0) &=& 0,
\label{eq:outer0-phi}\\
\partial_t u_0 &=& \nabla^2 u_0 + \frac{1}{2}\partial_t h(\phi_0).
\label{eq:outer0-u}
\end{eqnarray}
Introducing the function $f(\phi)$ into Eq.~(\ref{eq:outer0-phi}), 
we obtain $\phi_0=\pm 1$, and using that $h(\pm 1)=\pm 1$, 
Eq.~(\ref{eq:outer0-u}) adopts the form 
\begin{equation}
\partial_t u_0 = \nabla^2 u_0.
\end{equation}
\subsubsection{First Order}
At first order in $\varepsilon$, we find
\begin{eqnarray}
f''(\phi_0) \phi_1 &=& - \lambda g'(\phi_0) u_0,
\label{eq:outer1-phi}\\
\partial_t u_1 &=& \nabla^2 u_1+\frac{1}{2}\partial_t [\phi_1 h(\phi_0)].
\label{eq:outer1-u}
\end{eqnarray}
{}From Eq.~(\ref{eq:outer1-phi}) we determine $\phi_1=0$ by 
noting that the functions $f$, $g$ satisfy 
$f''(\pm 1) \neq 0$ and $g'(\pm 1) = 0$, and introducing 
$\phi_1=0$ into Eq.~(\ref{eq:outer1-u}) we get 
\begin{equation}
\partial_t u_1 = \nabla^2 u_1.
\end{equation}
\subsubsection{Second Order}
At second order ($\varepsilon^2$), and using that $\phi_1=0$, 
the random current ${\bf q}$ appears in the equation for the outer diffusive field
\begin{eqnarray}
f''(\phi_0)\phi_2  &=& -\lambda g'(\phi_0) u_1,
\label{eq:outer2-phi}\\
\partial_t u_2 &=& \nabla^2 u_2 + \frac{1}{2}\partial_t[h'(\phi_0)\phi_2] - {\bf \nabla}\cdot {\bf q}\;.
\label{eq:outer2-u}
\end{eqnarray}
Using that $g'(\pm 1)=0$ and $f''(\pm 1)\neq 0$, 
equation (\ref{eq:outer2-phi}) is solved by $\phi_2=0$, and the 
second term at the right hand side of Eq.~(\ref{eq:outer2-u}) can be neglected. 
Collecting the results obtained at the three first orders, 
the outer fields are given, up to second order in $\varepsilon$, by  
\begin{eqnarray}
\phi &=& \pm 1 + O(\varepsilon^3),
\label{eq:phi-diff}\\
\partial_{t} u &=& \nabla^{2}u -
{\bf \nabla}\cdot {\bf q}({\bf r},t) + O(\varepsilon^3)\;.
\label{eq:u-diff}
\end{eqnarray}
\subsection{Inner region}
\label{sec:b}
For the inner region, we write Eqs.~(\ref{eq:pf-1}), (\ref{eq:u-1})
in a curvilinear coordinate system centered at the interface. The idea is that the 
solvability condition for the very existence of solutions of these transformed
equations will provide the evolution of the coordinate system, {\it i.e.} of the
interface, which in fact constitutes the solution we are looking for.  To define
this coordinate system by maintaining it smooth at small scales, we use an auxiliary
coarse grained field defined as a local spatial and temporal average of the
fluctuating field $\phi$. The surface corresponding to the level set of this
coarse grained field $\langle \phi({\bf r},t) \rangle=0$ allows to define the
3D orthogonal curvilinear coordinate system 
$(r,s_1,s_2)$, where $r$ 
is a normal distance from the surface 
and $s_1$, $s_2$ are the arclength distances 
measured along the principal curvature directions of the surface. 
Furthermore, we introduce in the inner region the scaled 
normal coordinate $\rho=r/\varepsilon$ 
and the scaled time $\tau=t/\varepsilon$. 
We use capital letters to refer to all the fields when written in the inner region. 
After some manipulation, and keeping terms up to second order in 
$\varepsilon$, we obtain the inner equations in the frame of the moving interface 
%
\\
\begin{equation}
\begin{split}
&\alpha \varepsilon[\frac{d}{d\tau} - v \partial_\rho ]\Phi = 
\partial^2_\rho \Phi + \varepsilon \kappa \partial_\rho \Phi 
- \varepsilon^2 \rho (\kappa^2 -2 \Pi) \partial_\rho \Phi \\
&+ \varepsilon^2 \sum_{i=1,2} \partial^2_{s_i} \Phi 
- f'(\Phi) - \varepsilon \lambda  g'(\Phi) U 
+ \varepsilon^{1/2} H(\rho,{\bf s},\tau) , 
\label{eq:inner-phi_0}
\end{split}
\end{equation}
\begin{equation}
\begin{split}
&\frac{1}{\varepsilon}[\frac{d}{d\tau} - v\partial_\rho] U = 
\frac{1}{\varepsilon^2}\partial^2_\rho U 
+ \frac{1}{\varepsilon}[\kappa  -\varepsilon \rho (\kappa^2-2\Pi) ]\partial_\rho U \\
&+ \sum_{i=1,2} \partial^2_{s_i} U 
- \frac{v}{2 \varepsilon} \partial_\rho h(\Phi) 
+ \frac{1}{2 \varepsilon}\frac{d h(\Phi)}{d \tau} 
- \frac{1}{\varepsilon^2} \partial_\rho Q_\rho(\rho,{\bf s},\tau) ,
\label{eq:inner-u_0}
\end{split}
\end{equation}
where $v=v({\bf s},\tau)$ is the local normal velocity of the interface, and 
we have introduced $\kappa=\kappa_1+\kappa_2$ and $\Pi=\kappa_1 \kappa_2$ as the 
mean and Gaussian curvatures of the surface, being 
$\kappa_1({\bf s},\tau)$, $\kappa_2({\bf s},\tau)$ its principal curvatures. 

The fluctuating functions $H(\rho,{\bf s},\tau)= \varepsilon \eta({\bf r},t)$ and ${\bf Q}(\rho,{\bf s},\tau)= \varepsilon {\bf q}({\bf r},t)$ 
in Eqs.~(\ref{eq:inner-phi_0}), (\ref{eq:inner-u_0}) 
stand for the renormalized noises in the inner region, and $Q_\rho$ corresponds to the 
normal component of the stochastic current ${\bf Q}$. 
%
The correlations of these 
noise terms are given by
\begin{eqnarray}
\langle
H(\rho,{\bf s},\tau) H(\rho',{\bf s}',\tau') 
\rangle &=&
2 \sigma_\phi^2 
\delta(\rho-\rho')
\delta({\bf s}-{\bf s}') 
\delta(\tau-\tau') ,
\label{eq:eta-inner}\\
\langle
Q_i(\rho,{\bf s},\tau)
Q_j(\rho',{\bf s}',\tau')
\rangle &=&
2 \sigma_u^2 \delta_{ij} \delta(\rho-\rho') 
\delta({\bf s}-{\bf s}') 
\delta(\tau-\tau'),
\label{eq:q-inner}
\end{eqnarray}
so that the orders in $\varepsilon$ of $H(\rho,{\bf s},\tau)$ and ${\bf Q}(\rho,{\bf s},\tau)$ are those of $\sigma_\phi$ and $\sigma_u$ respectively.
Note that the renormalization of the noise terms
is a direct consequence of the 
scaling of the $t,r$ coordinates in the inner region. Indeed, noise 
terms give rise to an $\epsilon$ factor when written in the inner
region due to the rescaling in both time and normal distances 
of the delta functions $\delta(\rho) = \varepsilon\delta(r)$ 
and $\delta(\tau) = \varepsilon\delta(t)$.

Now we can see how the small noise assumption has been implemented in our approach.
With the choice given by Eqs.~(\ref{eq:sigma_order1}), (\ref{eq:sigma_order2})
for the orders in $\varepsilon$ of the noise amplitudes, 
$\varepsilon^{1/2} H$ is $O(\varepsilon^2)$ in Eq.~(\ref{eq:inner-phi_0}) 
and $-{\varepsilon^{-2}} \partial_\rho Q_\rho$ is $O(\varepsilon^0)$ in 
Eq.~(\ref{eq:inner-u_0}), {\it i.e.} one order higher than the temporal derivatives 
in these equations. In other words, both noise terms are first order 
perturbations for the dynamics in their respective inner equations.

At this point, we proceed as in the outer region by expanding the inner 
fields and parameters in powers of $\varepsilon$ 
\begin{eqnarray}
U &=& U_0+\varepsilon U_1 + \varepsilon^2 U_2 +O(\varepsilon^3),\\
\Phi &=& \Phi_0 + \varepsilon \Phi_1 + \varepsilon^2 \Phi_2 + O(\varepsilon^3),\\
\kappa_i &=& {\kappa_i}_0 + \varepsilon {\kappa_i}_1 + O(\varepsilon^2)\;,~i=1,2\\
\Pi &=& \Pi_0 + \varepsilon \Pi_1 + O(\varepsilon^2),\\
v &=& {v}_0 + \varepsilon {v}_1 +O(\varepsilon^2),
\label{eq:outera}
\end{eqnarray}
and inserting the expansions into the inner equations 
Eqs.~(\ref{eq:inner-phi_0}), (\ref{eq:inner-u_0}). The 
inner solutions will be obtained by matching with the 
outer solutions for $\rho \rightarrow \pm \infty$ and 
$r \rightarrow 0^\pm$, respectively. 
In the phase field equations, direct matching with the outer $\phi_i$ 
solutions Eq.~(\ref{eq:phi-diff}) provides the limiting boundary 
conditions for the $\Phi_i$ terms of the inner expansion  
\begin{eqnarray}
\Phi_0(\rho \rightarrow \pm \infty) &=& \mp 1, 
\label{eq:limiting0}\\
\Phi_i(\rho \rightarrow \pm \infty) &=& 0, ~\text{for} ~i=1,2.
\label{eq:limiting12}
\end{eqnarray}
Similarly, the matching condition for the inner diffusion 
field requires that, at leading order, the gradients of $U_0$ vanish
\begin{equation}
\lim_{\rho \rightarrow \pm \infty} \partial_\rho U_0 = 0.
\label{eq:mat0}
\end{equation}
At higher orders, the matching conditions for the diffusive field 
present some additional difficulties due to the apparition of 
random terms, and will be discussed in detail in Sec.~\ref{sec:d}.
\subsubsection{Zero order}
At leading order ($\varepsilon^0$ for the $\Phi$ equation, 
$\varepsilon^{-2}$ for the $U$ equation), the inner 
equations are given by 
\begin{eqnarray}
&&\partial^2_\rho \Phi_0 - f'(\Phi_0)=0,
\label{eq:inner-phi-0}\\
&&\partial^2_\rho U_0=0.
\label{eq:inner-u-0}
\end{eqnarray}
Inserting the double-well potential Eq.~(\ref{eq:double-well}) into 
Eq.~(\ref{eq:inner-phi-0}), we obtain the 
standard kink solution for the phase-field at zero-order
\begin{equation}
\Phi_0(\rho)=-\tanh\bigg(\frac{\rho}{\sqrt{2}}\bigg),
\label{eq:kinksol}
\end{equation}
which satisfies the matching condition Eq.~(\ref{eq:limiting0}) 
for $\rho \rightarrow \pm \infty$.
Integrating Eq.~(\ref{eq:inner-u-0}) twice over $\rho$, we have  
\begin{equation}
U_0(\rho,{\bf s},\tau)=A({\bf s},\tau) + B({\bf s},\tau) \rho,
\label{eq:U0-00}
\end{equation}
where $A$ and $B$ are integration constants. 
Imposing the matching condition Eq.~(\ref{eq:mat0}), we 
determine $B({\bf s},\tau)=0$ and obtain a $\rho$-independent 
solution for the diffusion field at zero order
\begin{equation}
U_0({\bf s},\tau)=A({\bf s},\tau).
\label{eq:U0-0}
\end{equation}
\subsubsection{First Order}
Using the solutions obtained at zero-order, the first-order inner 
equations ($\varepsilon^1$ for the $\Phi$ equation, 
$\varepsilon^{-1}$ for the $U$ equation) read
\begin{eqnarray}
\Omega \Phi_1 &=& 
-(v_0\alpha+\kappa_0)\partial_\rho \Phi_0 + \lambda g'(\Phi_0) U_0,
\label{eq:inner-phi-1}\\
\partial^2_\rho U_1 &=& \frac{d U_0}{d\tau} + \frac{v_0}{2}\partial_\rho h(\Phi_0),
\label{eq:inner-u-1}
\end{eqnarray}
where $\Omega$ is the self-adjoint operator 
$\Omega \equiv \partial^2_\rho- f''(\Phi_0)$ and we have used 
that $d\Phi_0/d\tau =0$ from Eq.~(\ref{eq:kinksol}). 
As described by Almgren \cite{almgren-one-sided}, 
an expression for $\Phi_1$ can be obtained from Eq.~(\ref{eq:inner-phi-1}) 
by inverting the operator $\Omega$, leading to 
\begin{equation} 
\Phi_1 = \Omega^{-1} [-(v_0\alpha+\kappa_0)\partial_\rho \Phi_0 + \lambda g'(\Phi_0) U_0].
\label{eq:Phi_1}
\end{equation}
Since $\Omega$ is an even operator and $\partial_\rho \Phi_0$, 
$g'(\Phi_0)$ are even functions of $\rho$, $\Phi_1$ 
is an even function of $\rho$.
Integrating Eq.~(\ref{eq:inner-u-1}) twice over $\rho$, we get 
\begin{equation}
U_1 = D({\bf s},\tau)+ C({\bf s},\tau) \rho +\frac{v_0}{2} \int_0^\rho d\rho' h(\Phi_0) 
+ \frac{1}{2}\frac{d U_0}{d\tau} \rho^2,
\label{eq:U1-0}
\end{equation}
where $D$ and $C$ are integration constants and 
we have used that $\partial_\rho U_0=0$ (cf. Eq.~(\ref{eq:U0-0})).
\subsubsection{Second order}
The stochastic terms appear in the inner equations 
at second order ($\varepsilon^2$ for the $\Phi$ equation, 
$\varepsilon^{0}$ for the $U$ equation), which are given by
\begin{equation}
\begin{split}
\Omega \Phi_2 = 
&-(\alpha v_1 + \kappa_1) \partial_\rho \Phi_0
-(\alpha v_0 + \kappa_0) \partial_\rho \Phi_1 
+ \alpha \frac{d \Phi_1}{d \tau}\\
&+\frac{1}{2}f'''(\Phi_0) \Phi_1^2 
+\rho (\kappa^2_0 -2 \Pi_0) \partial_\rho \Phi_0 \\
&+\lambda g'(\Phi_0) U_1 +\lambda g''(\Phi_0) \Phi_1 U_0 
- \varepsilon^{-3/2}H,
\label{eq:fi-o2}
\end{split}
\end{equation}
\begin{equation}
\begin{split}
\partial_\rho^2 U_2 
= &-(v_0+\kappa_0) \partial_\rho U_1 
+ \frac{d U_1}{d \tau} 
+ \frac{v_1}{2}\partial_\rho h(\Phi_0)
+ \frac{v_0}{2}\partial_\rho [h'(\Phi_0)\Phi_1]\\
&- \sum_{i=1,2} \partial^2_{s_i} U_0 - \frac{1}{2} h'(\Phi_0)\frac{d\Phi_1}{d\tau} 
+\frac{1}{\varepsilon^2} \partial_\rho Q_\rho
.
\label{eq:u-o2}
\end{split}
\end{equation}
The first equation Eq.~(\ref{eq:fi-o2}) will be used in the 
next section when imposing the second order solvability 
condition of the problem. Integrating Eq.~(\ref{eq:u-o2}) twice over $\rho$, we find
\begin{equation}
\begin{split}
&U_2 = F({\bf s},\tau) + E({\bf s},\tau)\rho -(v_0+\kappa_0) \int_0^\rho d\rho' U_1 \\
&+ \int_0^\rho d\rho' \int_0^{\rho'} d\rho'' \frac{d U_1}{d \tau} 
-\frac{1}{2}\partial^2_s U_0 \rho^2 
+\frac{v_1}{2}\int_0^\rho d\rho' h(\Phi_0) \\
&+\frac{v_0}{2} \int_0^\rho d\rho' h'(\Phi_0) \Phi_1
+ \frac{1}{\varepsilon^2} \int_0^\rho d\rho' Q_\rho(\rho,s,\tau),
\label{eq:U2-0}
\end{split}
\end{equation}
where $F$ and $E$ are again $\rho$-independent integration constants. 
%
%
%
\subsection{Solvability conditions}
\label{sec:c}
%
%
%
%
We impose now the solvability conditions for the inner problem, which at 
first and second orders are respectively given by
\begin{equation}
\int_{-\infty}^\infty (\partial_\rho \Phi_0) \Omega \Phi_j d\rho = 0, ~\text{for}~ j=1,2.
\label{eq:solvab}
\end{equation}
Inserting Eq.~(\ref{eq:inner-phi-1}) into the first order solvability 
condition, we get 
\begin{equation}
-(\alpha v_0 +\kappa_0) I_1 - \lambda I_2 U_0 = 0,
\end{equation}
which allows to determine $U_0$ as
\begin{equation}
U_0({\bf s},\tau) = - \frac{\alpha I_1}{\lambda I_2} v_0 - \frac{I_1}{\lambda I_2} \kappa_0,  
\label{eq:U0sol}
\end{equation}
where $I_1$ and $I_2$ are new integral constants given by
\begin{eqnarray}
I_1 &=& \int_{-\infty}^{\infty} d\rho  (\partial_\rho \Phi_0)^2,
\label{eq:I_1}\\
I_2 &=& -\int_{-\infty}^{\infty} d\rho g'(\Phi_0) (\partial_\rho \Phi_0).
\label{eq:I_2}
\end{eqnarray}
Imposing the second order solvability condition Eq.~(\ref{eq:solvab}), 
and taking into account the parity of the potentials $f,g,h$ and of the inner solutions 
$\Phi_0,\Phi_1$, we determine an expression for the constant $D$ in Eq.~(\ref{eq:U1-0})
%
%
%
%
\begin{equation}
D({\bf s},\tau)= - (\alpha v_1 + \kappa_1)\frac{I_1}{\lambda I_2} +  v_0\frac{I_3}{2 I_2} + \frac{I_4}{2 I_2} + 
\frac{\alpha I_5}{\lambda I_2} - \varepsilon^{-3/2}\frac{Z({\bf s},\tau)}{\lambda I_2},
\label{eq:D}
\end{equation}
where $I_3$, $I_4$ and $I_5$ are defined by 
\begin{eqnarray}
I_3 &=& \int_{-\infty}^{\infty} d\rho \; (\partial_\rho \Phi_0) \; g'(\Phi_0)
\int_{0}^{\rho} d\rho' h(\Phi_0),
\label{eq:I_3}\\
I_4 &=&  \frac{d U_0}{d\tau} 
\int_{-\infty}^{\infty} d\rho \; (\partial_\rho \Phi_0) 
g'(\Phi_0) \rho^2,
\label{eq:I_4}\\
I_5 &=& \int_{-\infty}^{\infty} d\rho \; (\partial_\rho \Phi_0)\frac{d \Phi_1}{d\tau},
\label{eq:I_5}
\end{eqnarray}
and $Z$ is a stochastic term given by
\begin{equation}
Z({\bf s},\tau) = \int_{-\infty}^{\infty} d\rho \;(\partial_\rho \Phi_0)\; 
H(\rho,s,\tau),
\label{eq:Z}
\end{equation}
whose statistical properties are given by
\begin{equation}
\langle 
Z({\bf s},\tau) Z({\bf s}',\tau') 
\rangle = 
2 I_1 \sigma_\phi^2 \delta({\bf s}-{\bf s}') \delta(\tau-\tau').
\label{eq:Z-corra}
\end{equation}

\subsection{Matching of fluctuating fields}
\label{sec:d}
At this point, we continue by imposing the remaining 
asymptotic matching conditions of the problem. 
However, the matching of the diffusion field presents some 
subtleties due to its fluctuating character.
The main problem is that, at second order in $\varepsilon$, 
the $U$ field fluctuates in the normal direction (cf. Eq.~(\ref{eq:U2-0})),
and hence cannot be written as a simple asymptotic expansion for 
$\rho \rightarrow \pm \infty$, preventing the matching with the outer field. 
This difficulty can be overcome by introducing an auxiliary matching function 
defined in both regions as
\begin{eqnarray}
\chi (r,{\bf s},t) &=& u(r,{\bf s},t) - \int_0^r dr' q_r(r',{\bf s}',t),
\label{eq:outer-chi} \\
X(\rho,{\bf s},\tau) &=& U(\rho,{\bf s},\tau) 
- \int_0^\rho d\rho' Q_\rho (\rho',{\bf s},\tau). 
\label{eq:X-chi}
\end{eqnarray}
In view of Eq.~(\ref{eq:U2-0}), it is easy to see that the inner 
auxiliary function $X$ introduced in Eq.~(\ref{eq:X-chi}) is smooth up to 
order $\varepsilon^2$ in the matching region $r_M$.
%
%
Explicitely, if $X$ is asymptotically 
expanded for $\rho \rightarrow \pm \infty$ as 
\begin{equation}
X \sim T + S\; \rho + R \; \rho^2 + O(\rho^3),
\label{eq:Xexp}
\end{equation}
and the outer matching function $\chi$ is expanded in Taylor around $r=0^{\pm}$ by 
\begin{equation}
\chi \approx \chi(0^\pm)+\partial_r \chi(0^\pm) \cdot r 
+ \frac{1}{2}\partial^2_r \chi(0^\pm) \cdot r^2+ O(r^3),
\label{eq:xexp}
\end{equation}
the inner and outer terms can be matched at $r_M \sim \varepsilon^{1/2}$ 
in the limit $\varepsilon \rightarrow 0$ to obtain the matching relations 
\begin{eqnarray}
T &=& \chi(0^\pm),
\label{eq:matching-1}\\
S &=&\varepsilon \partial_r \chi(0^\pm),
\label{eq:matching-2}\\
R &=&\frac{\varepsilon^2}{2} \partial^2_r \chi(0^\pm).
\label{eq:matching-3}
\end{eqnarray}
The last step is to expand the previous 
Equations~(\ref{eq:matching-1}), (\ref{eq:matching-2}) and (\ref{eq:matching-3}) 
in powers of $\varepsilon$ to complete
the matching at each order in $\varepsilon$.

\subsubsection{First Order}
At first order in $\varepsilon$, the inner field $U_1$ given by Eq.~(\ref{eq:U1-0}) 
can be asymptotically expanded for $\rho \rightarrow \pm \infty$ as 
\begin{equation}
U_1 \sim \frac{1}{2}\frac{d U_0}{d\tau} \rho^2 + \bigg[ C \mp \frac{v_0}{2}\bigg]\rho + 
D + \frac{v_0}{2}J_1^\pm,
\label{eq:U_1exp}
\end{equation}
where $D$ is given by Eq.~(\ref{eq:D}) and 
\begin{equation}
J_1^\pm = \int_0^{\pm \infty} d\rho [h(\Phi_0(\rho))\pm 1 ],
\end{equation}
where we have used that $h$ satisfies $h(\pm 1) =\pm 1$ and the far field condition 
Eq.~(\ref{eq:limiting0}). Since $h(\Phi_0)$ is an odd function of $\rho$, we have 
\begin{equation}
J^+_1 = J^-_1 \equiv J_1.
\end{equation}
\subsubsection{Second Order}
Similarly, the second order inner solution for the diffusive field 
Eq.~(\ref{eq:U2-0}) can be expanded asymptotically for $\rho \rightarrow \pm \infty$ as
\begin{equation}
\begin{split}
&U_2 \sim F + E\rho 
-(v_0+\kappa_0)\int_0^\rho d\rho' \frac{d U_1}{d\tau} 
+ \int_0^\rho d\rho' \int_0^{\rho'} d\rho'' \frac{d U_1}{d\tau} \\
&-\frac{1}{2}\partial^2_s U_0 \rho^2 
-\int_0^\rho d\rho' \int_0^{\rho'} d\rho'' h'(\Phi_0)\frac{d\Phi_1}{d\tau} \\
& + \frac{v_0}{2}J_2^\pm + \frac{v_1}{2}J_1 \mp \frac{v_1}{2}\rho
+\frac{1}{\varepsilon^2}\int_0^\rho d\rho' Q_\rho(\rho,{\bf s},t),
\label{eq:U_2exp}
\end{split}
\end{equation}
%
where
\begin{equation}
J_2^\pm = \int_0^\rho d\rho' h'(\Phi_0)\Phi_1,
\end{equation}
and we have used the far field conditions 
Eqs.~(\ref{eq:limiting0}) and (\ref{eq:limiting12}).
Inserting the expressions Eqs.~(\ref{eq:U0sol}), (\ref{eq:U_1exp}) 
and (\ref{eq:U_2exp}) into the right hand side of Eq.~(\ref{eq:X-chi}), we can 
determine the parameters $R,S$ and $T$ in the far field expansion 
of the matching function $X$ (cf. Eq.~(\ref{eq:Xexp})) and 
perform the matching with the outer function $\chi$ (cf. Eq.~(\ref{eq:xexp})). 

Imposing the third matching condition Eq.~(\ref{eq:matching-3}) at first order 
in $\varepsilon$, we determine that 
\begin{equation}
\frac{dU_0}{d\tau}=0,
\end{equation}
which, using the relation Eq.~(\ref{eq:Phi_1}), brings to 
\begin{equation}
\frac{d\Phi_1}{d\tau}=0,
\end{equation}
and therefore the integral constants $I_4$ and $I_5$ defined in Eqs.~(\ref{eq:I_4}), 
(\ref{eq:I_5}) vanish
\begin{equation}
I_4=I_5=0.
\end{equation}
%
{}From the two first orders of Eq.~(\ref{eq:matching-1}), 
we get an expression for the outer diffusive field at the interface valid 
up to first order
\begin{equation}
u(0^\pm) = - \frac{I_1}{\lambda I_2} (\alpha v + \kappa) +
\frac{\varepsilon v_0}{2}(\frac{I_3}{I_2} + J_1) 
- \frac{z({\bf s},t)}{\lambda I_2} 
+ O(\varepsilon^2),
\label{eq:GT-PF}
\end{equation}
where $z({\bf s},t)= Z({\bf s},\tau)\varepsilon^{-1/2}$ is a stochastic 
term whose statistical properties can be determined from Eq.~(\ref{eq:Z-corra}) 
and are given by
\begin{equation}
\langle 
z({\bf s},t) z({\bf s}',t') 
\rangle = 
2 I_1 \sigma_\phi^2 \delta({\bf s}-{\bf s}') \delta(t-t').
\label{eq:Z-corr}
\end{equation}
Note that the projected interfacial noise term has neither in Eq.~(\ref{eq:GT-PF}) nor in Eq.~(\ref{eq:Z-corr}) 
any explicit dependence in $\varepsilon$, which
is a direct consequence of the $\varepsilon^{3/2}$ factor introduced in the noise 
term of Eq.~(\ref{eq:pf-1}). Indeed, this is the reason why such factor was introduced 
in the formulation of the model.

The calculation is completed by imposing the matching condition
Eq.~(\ref{eq:matching-2}) up to second order, which can be written as
\begin{equation}
v_0 + \varepsilon v_1 = \partial_r \chi|^-_+ + O(\varepsilon^2).
\label{eq:cons_1}
\end{equation}
Inserting Eq.~(\ref{eq:outer-chi}) into Eq.~(\ref{eq:cons_1}), we get a heat/mass 
conservation equation valid up to first order in $\varepsilon$,
\begin{equation}
v = v_0 + \varepsilon v_1 = [\partial_r u]^-_+ - [q_r]^-_+ + O(\varepsilon^2),
\label{eq:pf-cons}
\end{equation}
where $q_r$ accounts for a normal stochastic current across the interface. 
This term, although being of order $\varepsilon^2$, has not been neglected 
in Eq.~(\ref{eq:pf-cons}) in order to not break mass conservation in the 
stochastic diffusion equation Eq.~(\ref{eq:u-diff}), which is valid up to second order. 

This last equation completes 
the sharp-interface projection of the stochastic phase-field model of
Eqs.~(\ref{eq:pf-1})-(\ref{eq:q}). 
This projection constitutes the main result of this paper, and is given by the
diffusion equation Eq.~(\ref{eq:u-diff}), with the noise of Eq.~(\ref{eq:q}), 
supplemented with two moving boundary conditions at the
interface: the conservation condition Eq.~(\ref{eq:pf-cons}), and
the Gibbs-Thomson Eq.~(\ref{eq:GT-PF}), where a new projected 
interfacial noise appears with correlation given by Eq.~(\ref{eq:Z-corr}). 
Note that the projected boundary conditions at the interface Eqs.~(\ref{eq:pf-cons}), 
(\ref{eq:GT-PF}) are obtained at the order immediately lower than the order at which 
the asymptotic expansion is performed. 

While the general lines of the calculation follow 
the standard sharp-interface asymptotics, we have included the fluctuation 
terms during all the procedure, which have been projected in the weak noise
limit. This calculation is thus similar to the front
dynamics projection performed in Ref.~\cite{andrea-laure}. 
Indeed, the projected interfacial noise appearing
in Eq.~(\ref{eq:GT-PF}) is the analogous counterpart of the noise term of the
projected eikonal front equation of Ref.~\cite{andrea-laure}.
%

\section{Internal fluctuations in a generic phase field model}
\label{sec:parameters}
Thus far, the noises considered in this work are intended 
to account for both external and
internal sources of fluctuations. Nevertheless, it is worth 
pointing out that the resulting stochastic sharp-interface equations
are similar to those postulated in the Langevin formulation of solidification due
to Karma \cite{karma-fluct-prl,karma-fluctuations} 
(see also Ref.~\cite{cherepanova}), which was constructed to follow 
equilibrium statistics. This offers the possibility of using the results above to
provide generic (not necessarily variational) phase-field models with the correct
equilibrium fluctuations. 
To illustrate this, let us consider the Langevin sharp-interface 
equations \cite{karma-fluct-prl,karma-fluctuations} 
\begin{eqnarray}
\partial_t u_{SI} &=& \nabla^2 u_{SI} - {\bf \nabla} \cdot {\bf q}^{SI}({\bf r},t),
\label{eq:si-diff}\\
v_{SI} &=& [\partial_r u_{SI}]^-_+ - [q_r^{SI}]^-_+, 
\label{eq:si-cons}\\
u_{SI}(0) &=& -d_0 \kappa - \beta v + \theta({\bf r},t)\;,
\label{eq:si-GT}
\end{eqnarray}
where ${\bf q}^{SI}$ and $\theta$ are fluctuating terms with correlations 
given by
\begin{eqnarray}
\langle q^{SI}_i({\bf r},t) q^{SI}_j({\bf r}',t') \rangle &=&
\frac{2K_BT_M^2c}{L^2 l^d} \delta_{ij}  \delta({\bf r}-{\bf r'}) 
\delta(t-t'),
\label{eq:si-q-fluct}\\
\langle \theta({\bf s},t) \theta({\bf s}',t') \rangle &=& 
\frac{2K_BT_M^2 c \beta}{L^2 l^d} \delta({\bf s}-{\bf s'}) \delta(t-t'). 
\label{eq:si-fi-fluct} 
\end{eqnarray}
The Gibbs-Thompson equation Eq.~(\ref{eq:si-GT}) can be compared 
with Eq.~(\ref{eq:GT-PF}) and the diffusion equation 
Eq.~(\ref{eq:si-diff}) with Eq.~(\ref{eq:u-diff}).
This comparison enables the determination of the 
phase-field parameters in terms of physical 
and substance parameters, which are given by the equations
\begin{eqnarray}
\lambda &=& \frac{I_1}{I_2 d_0}, 
\label{eq:lambdaa}\\
\alpha &=& \frac{\beta}{d_0},
\label{eq:alphaa}\\
\sigma_u^2 &=& \frac{K_BT_M^2c}{L^2 l^d},
\label{eq:sigmauu}\\
\sigma_\phi^2 &=& \frac{I_1 K_BT_M^2 \; c \; \beta}{d_0^2L^2 l^d}.
\label{eq:sigmaphii}
\end{eqnarray}
%
%
In the last relations, the two first equations Eqs.~(\ref{eq:lambdaa}), 
(\ref{eq:alphaa}) are the usual expressions determined by the standard 
asymptotic procedure, whereas Eq.~(\ref{eq:sigmauu}) reflects the identification between the conserved stochastic currents of both phase field model and sharp interface projection.
In this sense, a major result of 
our approach has been the derivation of an expression for the noise strength 
of the phase-field, that is Eq.~(\ref{eq:sigmaphii}), from the above calculations.

With this election of the model parameters, the phase-field simulations 
will present the correct equilibrium statistics in the limit of small interface 
thickness $\varepsilon \rightarrow 0$. Therefore, the non-variational 
phase-field formulation of Eqs.~(\ref{eq:pf-1}), (\ref{eq:u-1}) can be used 
to quantitatively account for thermodynamical fluctuations in solidification 
processes. 

\section{TEST OF THE APPROACH}
\label{sec:test}
In order to test the validity of our approach, 
we have performed 2D phase-field simulations to obtain the 
power spectrum of the interfacial fluctuations of a 
solid-liquid stationary flat interface. 
Introducing the Fourier transform 
of the interface position $\xi({\bf r},t)$ as 
$\xi_{k}(t) = \int dk \; \xi({\bf r},t) \; e^{-i {\bf k} {\bf r}}$, 
the power spectrum of a stationary planar front 
in scaled variables is given by
\begin{equation}
S(k)=\langle \xi_k \xi_{-k} \rangle = 
\int \frac{dk'}{2\pi} \langle \xi_k \xi_{k'} \rangle=
\frac{K_B T_M}{\gamma}\frac{1}{k^2},
\label{eq:power1}
\end{equation}
where $\gamma=l^d L^2 d_0/T_M c$ is 
the scaled interfacial surface energy. 
In the simulations, space and times have been scaled 
using length and time scales of $l=10^{-8} \; m$ and 
$\nu=9 \times 10^{-10} \; s$, respectively.
The functions $h,g$ have been chosen to be 
$h(\phi)=\phi$ and $g'(\phi)=(1-\phi^2)^2$ 
so that the model does not have a variational 
structure. The substance parameters used in the simulations 
correspond to the values of the pure SCN in the 3D case, 
and are given by $d_0=0.2817$, $\beta=3.0331$ \cite{chernov,aziz_comment} 
and $\sigma^2_u=0.001432$. 
For this choice, and using Eqs.~(\ref{eq:lambdaa})-(\ref{eq:sigmaphii}), 
the phase-field parameters take the values 
$\lambda=3.13$, $\alpha=10.76$ and $\sigma^2_{\phi}=0.05158$. 
The interface thickness has been taken to be $\varepsilon=0.3$. 

The simulations have been implemented with a finite differences scheme on 
a $50 \times 512$ lattice with $\Delta x=\Delta y=0.2$ and $\Delta t=0.005$.
We have used the initial conditions 
$\phi(x,y,0)=-\tanh(x/\varepsilon \sqrt{2})$, $u(x,y,0)=0$. 
Non-flux and periodic boundary conditions 
have been imposed in the $x$ and $y$ directions respectively. 

The numerical implementation of the stochastic terms has been 
carried out by generating Gaussian-distributed random numbers 
at each of the lattice sites. The correlations of these numbers can be 
determined by discretizing the time and spatial delta functions in 
Eqs.~(\ref{eq:eta}), (\ref{eq:q}) by substituting 
$\delta(x-x') \rightarrow \delta_{ii'}/\Delta x$ and 
$\delta(t-t') \rightarrow \delta_{nn'}/\Delta t$.
The divergence of the stochastic current in Eq.~(\ref{eq:u-1}) 
has been discretized by using a forward differences scheme 
${\bf \nabla}\cdot {\bf q}({\bf r},t)|_{i,j}= 
(q_x(i+1,j)- q_x(i,j))/\Delta x + 
(q_y(i,j+1)-q_y(i,j))/\Delta y$.

\begin{figure}
\vspace{0.15cm}
\includegraphics[width=0.5\textwidth]{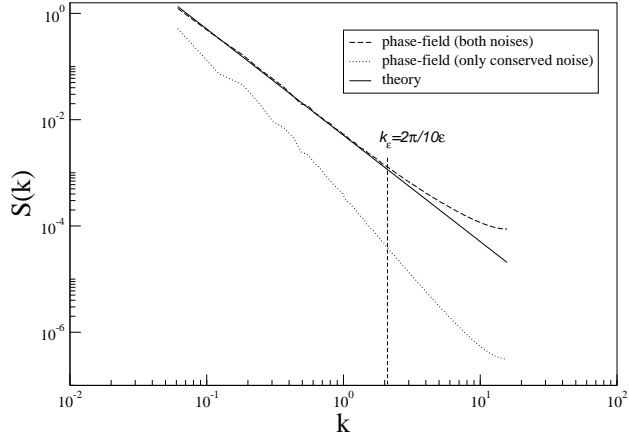}
\caption{Comparison between the theoretical power 
spectrum of the stationary interface and the 
results from the phase-field simulations.}
\label{fig:fig1}
\end{figure}

The power spectrum statistics has been 
obtained as a time average among the last $3\times10^6$ time
steps in a long-term simulation of $3.5\times10^6$ steps, and 
is represented by a dashed line in Fig.~\ref{fig:fig1}.   
The solid line in Fig.~\ref{fig:fig1} depicts the theoretical 
prediction given by Eq.~(\ref{eq:power1}) and, as it can be seen, 
an excellent agreement is found between theoretical and numerical results. 
The vertical dashed line in the figure represents the wavelength associated with the 
effective thickness of the interface, and determines the expected breakdown 
of the phase-field description. 

\section{DISCUSSION AND CONCLUSIONS}
\label{sec:discuss}

To summarize, we have obtained an asymptotic projection of the 
fluctuating phase-field equations (\ref{eq:pf-1}), (\ref{eq:u-1}) 
to a sharp interface description. 
This has been worked out by means of a hybrid asymptotic procedure, 
combining sharp interface and small noise limits. 
As a result, the projected equations adopt the form of a 
moving boundary problem with a conserved stochastic 
force ${\bf q}$ in the equations for the diffusion field 
Eqs.~(\ref{eq:u-diff}) and (\ref{eq:pf-cons}), 
and an interfacial noise $z({\bf s},t)$ in the Gibbs-Thompson 
condition Eq.~(\ref{eq:GT-PF}). 
Other authors have previously introduced fluctuations in phase-field 
models \cite{karma-pf-noise,pavlik-sekerka-1,pavlik-sekerka-2}, 
but their approaches applied only for the case 
of variational formulations and were restricted to noises from a 
thermodynamical origin. 

In this context, it has been claimed \cite{karma-pf-noise} that the 
presence of a non-conserved noise such as $\eta$
in the equation for the phase field
is not relevant for the dynamics of the phase field model, 
and thus could be omitted in simulations. 
In order to check the importance of the non-conserved phase field noise,  
we have carried out a numerical test with the same parameters reported in 
section ~\ref{sec:test} but taking $\sigma_\phi=0$. In this case, 
the power spectrum is plotted as a dotted line in Fig.~\ref{fig:fig1}. 
The clear disagreement with both the theoretical prediction and the simulations of the complete model indicates that the phase field noise is indeed necessary 
in order to obtain quantitative results. 
This can be explained by noting that Eq. (\ref{eq:GT-PF}) establishes 
a direct relation between the non-conserved phase-field noise $\eta$ and 
the interfacial fluctuations appearing in the Gibbs-Thompson equation, usually 
associated to kinetic attachment effects \cite{karma-fluctuations}. 
Thus, the small significance of the $\eta$ noise reported in 
Ref.~\cite{karma-pf-noise} is probably due to the fact that
the stationary power spectrum was calculated in the limit 
of vanishing kinetics $\beta=0$. 
Therefore, we conclude that, in the presence of kinetic effects, the phase field 
noise is relevant for a quantitative description of the solidification process. 

It is interesting to discuss
the scaling of the noise terms as proposed on the one hand 
in Eqs.~(\ref{eq:sigma_order1}), (\ref{eq:sigma_order2}) and on the other hand in the $\varepsilon^{3/2}$ factor explicitly appearing in the equation for the phase field, Eq.~(\ref{eq:pf-1}). 
As it has already been commented, the assumption of the order relations
Eqs.~(\ref{eq:sigma_order1}), (\ref{eq:sigma_order2}) has permitted to manage a double expansion (sharp interface and small noise) by formally using a single small parameter. The specific powers of $\varepsilon$ appearing in these relations have been chosen for maintaining fluctuations as small perturbations for the dynamics of both 
inner and outer equations.
On the contrary, the multiplicative factor of the noise term of the equation for the phase field, Eq.~(\ref{eq:pf-1}), has a different motivation.
It is well known that there are problems in the formulation 
of stochastic field equations when the noise terms are 
delta-correlated in space. I such cases, some kind of regularization 
is required. In Ref.~\cite{andrea-laure}, for instance, this 
regularization was provided by the correlation 
length of the noise, in such a way that the results did
depend on that parameter. In the present case, the regularization is
provided by the interface width $\varepsilon$. 
The projection of the bulk noise into the interface gives a 
fluctuation term that in principle should diverge as 
$\epsilon$ goes to zero. The $\epsilon^{3/2}$ factor of the noise 
term in Eq.~(\ref{eq:pf-1}) exactly cancels out this divergence, and has been 
introduced in the formulation of the model precisely to make 
results independent of $\epsilon$, specifically regarding the new interfacial noise term $z$ in Eqs.~(\ref{eq:GT-PF}), (\ref{eq:Z-corr}). 

In conclusion, we have proposed an asymptotic procedure  
to obtain the sharp-interface projection of a generic 
(not necessarily variational) 
phase-field model with fluctuations. We have tested the validity 
of our approach by comparing the  
phase-field results with the theoretical prediction for the 
interfacial fluctuations in a simple solidification problem. 
This procedure can be useful both in situations where the 
fluctuation-dissipation theorem does not hold, such as in  
presence of external fluctuations, and when the phase-field model has a 
non-variational nature. An important example of this latter 
case corresponds to efficient phase-field models for the 
solidification of alloys \cite{karma-thin,karma-one-sided}, which with our method can 
incorporate internal fluctuations. 
The availability of these models 
for quantitative simulations in the fluctuating case appears as a 
promising step towards the study of complex situations such as  
the apparition of dendritic sidebranching or the 
wavelength selection during initial redistribution transients in  
the directional solidification of alloys \cite{benitezFNL,benitezfuture}. 

\section{ACKNOWLEDGMENTS}
This work was financially supported by
Direcci\'on General de Investigaci\'on Cient\'{\i}fica y T\'ecnica
(Spain) (Project BFM2003-07850-C03-02) and
Comissionat per a Universitats i Recerca (Spain) (Project
2001/SGR/00221). 
We also acknowledge computing
support from Fundaci\'o Catalana per a la Recerca through C4 and CESCA (Spain).
\bibliography{ben_pre}         
\end{document}